\documentclass[11pt]{article}
\usepackage{cleveref}
\usepackage{latexsym}
\newtheorem{theorem}{Theorem}
\newtheorem{lemma}{Lemma}
\newcommand{\dfn}{\stackrel{\triangle}{=}}
\newcommand {\exe} {\stackrel{\cdot} {=}}
\newcommand {\gexe} {\stackrel{\cdot} {\ge}}
\newcommand {\lexe} {\stackrel{\cdot} {\le}}

\newcommand {\hH} {\hat{H}}
\newcommand {\hP} {\hat{P}}

\newcommand {\bx} {\mbox{\boldmath $x$}}

\newcommand {\by} {\mbox{\boldmath $y$}}
\newcommand {\bz} {\mbox{\boldmath $z$}}

\newcommand {\bE} {\mbox{\boldmath $E$}}

\newcommand {\bX} {\mbox{\boldmath $X$}}
\newcommand {\bY} {\mbox{\boldmath $Y$}}

\newcommand{\calG}{{\cal G}}

\newcommand{\calL}{{\cal L}}

\newcommand{\calP}{{\cal P}}

\newcommand{\calS}{{\cal S}}
\newcommand{\calT}{{\cal T}}

\newcommand{\calX}{{\cal X}}
\newcommand{\calY}{{\cal Y}}
\newcommand{\calZ}{{\cal Z}}
\begin{document}
\thispagestyle{empty}
\setcounter{page}{1}
\setlength{\baselineskip}{1.5\baselineskip}

\title{Universal Randomized Guessing
with Application to Asynchronous Decentralized Brute--Force Attacks}
\author{Neri Merhav\thanks{N. Merhav is with the Andrew and Erna Viterbi Faculty of Electrical Engineering, Technion -- Israel Institute of Technology, Haifa 32000, Israel. E-mail: {\tt merhav@technion.ac.il}} 
\and
Asaf Cohen\thanks{A. Cohen is with the Department of Communication System Engineering, Ben Gurion University of the Negev, Beer Sheva 84105, Israel. E-mail: {\tt coasaf@bgu.ac.il}}}
\maketitle

\begin{abstract}
Consider the problem of guessing the realization of a random vector $\bX$ by repeatedly submitting queries (guesses) of the form ``Is $\bX$ equal to $\bx$?" until an affirmative answer is obtained.
In this setup, a key figure of merit is the number of queries required until the right vector is identified, a number that is termed the \emph{guesswork}. Typically, one wishes to devise a guessing strategy which minimizes a certain guesswork moment.

In this work, we study a universal, decentralized scenario where the guesser does not know the distribution of $\bX$, and is not allowed to use a strategy which prepares a list of words to be guessed in advance, or even remember which words were already used. Such a scenario is useful, for example, if bots within a Botnet carry out a brute--force attack in order to guess a password or decrypt a message, yet cannot coordinate the guesses between them or even know how many bots actually participate in the attack.

We devise universal decentralized guessing strategies, first, for memoryless sources, and then generalize them for finite--state sources. In each case, we derive the guessing exponent, and then prove its asymptotic optimality by deriving  a compatible converse bound. The strategies are based on randomized guessing using a universal distribution. We also extend the results to guessing with side information. Finally, for all above scenarios, we design efficient algorithms in order to sample from the universal distributions, resulting in strategies which do not depend on the source distribution, are efficient to implement, and can be used asynchronously by multiple agents.  
$$
$$
\end{abstract}
\clearpage
\section{Introduction}
Consider the problem of guessing the realization of a random $n$--vector $\bX$ using a sequence of yes/no queries of the form: ``Is $\bX=\bx_1$?'', ``Is $\bX=\bx_2$?'' and so on, until an affirmative answer is obtained. Given a distribution on $\bX$, a basic figure of merit in such a guessing game is the \emph{guesswork}, defined as the number of trials required until guessing the right vector. 

Devising \emph{guessing strategies} to minimize the guesswork, and obtaining a handle on key analytic aspects of it, such as its \emph{moments} or its \emph{large deviations} rate function, has numerous applications in information theory and beyond. For example, sequential decoding \cite{wozencraft1957sequential,Arikan96} or guessing a codeword which satisfies certain constraints \cite{PfisterSullivan}. In fact, since the ordering of all sequences of length $n$ in a descending order of probabilities is, as expected, the optimal strategy under many optimality criteria\footnote{Specifically, if $G(\bX)$ is the guesswork, this order minimizes $\bE\{F[G(\bX)]\}$ for any monotone non-decreasing function $F$.}, the guessing problem is intimately related to fixed-to-variable source coding without the prefix constraint, or \emph{one-shot coding},
where it is clear that one wishes to order the possible sequences in a descending probability of appearance before assigning them codewords \cite{Szpankowski11F2V,Kontoyiannis14optimal_lossless,Kosut_universal_F2V_17}.

Contemporary applications of guesswork focus on information security, that is, guessing passwords or decrypting messages protected by random keys. E.g., one may use guessing strategies and their guesswork exponents while proactively trying to crack passwords, as a mean of assessing password security within an organization \cite{bishop1995improving,dell2010password}. Indeed, it is increasingly important to be able to assess password strength \cite{kelley2012guess}, especially under complex (e.g., non-i.i.d.) password composition requirements. While the literature includes several studies assessing strength by measuring how hard it is for common cracking methods to break a certain set of passwords \cite{dell2010password,kelley2012guess} or by estimating the entropy of passwords created under certain rules \cite{komanduri2011passwords}, the guesswork remains a key \emph{analytic tool} in assessing password strength for a given sequence length and distribution. As stated in \cite{bonneau2012science}, ``we are yet to see compelling evidence that motivated users can choose passwords which resist guessing by a capable attacker". Thus, analyzing the guesswork is useful in assessing how strong a key--generation system is, how hard will it be for a malicious party to break it, or, from the malicious side point of view, how better is one guessing strategy compared to the other. 

Arguably, human--created passwords may be of a finite, relatively small length, rather than long sequences which justify asymptotic analysis of the guesswork. Yet, as mentioned above, the guesswork, as a key figure of merit, may be used to aid in assessing computer generated keys  \cite{pliam00cryptology} or passwords as well. For example, a random key might be of tens or even hundreds of bits long, and passwords saved on servers are often \emph{salted} before being hashed, resulting in increased length \cite{Gauravaram12salt}. Moreover, experiments done on finite block lengths agree with the insights gained from the asymptotic analysis \cite{rezaee17budget}. As a result, large deviations and asymptotic analysis remain as key analytic tools in assessing password strength \cite{christiansen2013guessing,christiansen2013guesswork,rezaee17budget,yona17bias,salamatian2017centralized}. Such asymptotic analysis provides us, via tractable expressions, the means to understand the guesswork behaviour, the effect various problem parameters have on its value, and the fundamental information measures which govern it. E.g., while the entropy is indeed a  relevant measure for ``randomness" in passwords \cite{komanduri2011passwords}, via asymptotic analysis of the guesswork \cite{Arikan96} we now know that the R\'enyi entropy is the right measure when \emph{guessing} or even a distributed brute--force attack \cite{salamatian2017centralized,SHBCM18}. Non--asymptotic results, such as the converse result in  \cite{courtade14}, then give us finer understanding of the dependence on the sequence length.

Keeping the above applications in mind, it is clear the vanilla model of a single, all-capable attacker, guessing a password $\bX$ drawn from an i.i.d.\ source of a known distribution, is rarely the case of interest. In practical scenarios, several intricacies complicate the problem. While optimal passwords should have maximum entropy, namely, be memoryless and uniformly distributed over the alphabet, human-created passwords are hardly ever such. They tend to have memory and  a non--uniform distribution \cite{malone2012investigating}, due to the need to remember them as well as many other practical considerations (e.g., keyboard structure or the native language of the user) \cite{1341406,Vishwakarma14dictionary}. Thus, the ability to efficiently guess non-memoryless passwords and analyze the performance of such guessing strategies is crucial. 

Moreover, the underlying true distribution is also rarely known. In \cite{malone2012investigating}, the authors investigated the distribution of passwords from four known databases, and tried to fit a Zipf distribution\footnote{In this model, the probability of password with rank
$i$ is $P_i = K i^{−s}$, where $s$ is a parameter and $K$ is a normalizing constant}. While there was no clear match, it was clear that a small parameter $s$ is required, to account for a heavy tail. Naturally, \cite{malone2012investigating} also stated that ``If the right distribution of passwords can be identified, the cost of guessing a password can be reduced". 

Last but not least, from the attacker's side, there might be additional information which facilitates the guessing procedure on the one hand, yet there might be restrictions that prevent him/her from carrying out the optimal guessing strategy. That is, on the one hand, the attacker might have \emph{side information}, e.g., passwords for other services which are correlated with the one currently attacked, and thereby significantly decrease the guesswork \cite{Arikan96,4036408,christiansen2013guessing,salamatian2017centralized}. On the other hand, most modern systems will limit the ability of an attacker to submit too many queries from a single IP address, hence to still submit a large amount, these must be submitted from different machines. Such machines may not be synchronized\footnote{When bots or processes working in parallel are able to be completely synchronized, they may use a pre-compiled list of usernames and passwords - ``hard-coding" the guessing strategy \cite{owens2008study,tirado2018new}.}, namely, one may not know which queries were already submitted by the other. Moreover, storing a large (usually, an exponentially large) list of queries to be guessed might be a too heavy burden, especially for small bots in the botnet (e.g., IoT devices). The attacker is thus restricted to \emph{distributed brute force attacks}, where numerous devices send their queries simultaneously, yet without the ability to synchronize, without knowing which queries were already sent, or which bots are currently active and which ones failed \cite{SHBCM18}.
\subsection*{Main Contributions}

In this paper, we devise \emph{universal}, \emph{randomized} (hence, decentralized) guessing strategies for a wide family of information sources, and assess their performance by analyzing their guessing moments, as well as exponentially matching converse bounds, thereby proving their asymptotic optimality.

Specifically, we begin from the class of memoryless sources, and propose a guessing strategy. The strategy is universal both in the underlying source distribution and in the guesswork moment to be optimized. It is based on a \emph{randomized} approach to guessing, as opposed to an ordered list of guesses, and thus it can be used by asynchronous agents that submit their guesses concurrently. We prove that it achieves the optimal guesswork exponent, we provide an efficient implementation for the random selection of guesses, and finally, extend the results to guessing with side information.

Next, we broaden the scope to a wider family of non--unifilar finite--state sources, namely, hidden Markov sources. We begin with a general converse theorem and then provide a simple matching direct theorem, based on deterministic guessing. We then provide an alternative direct theorem, that employs a randomized strategy, which builds on the Lempel--Ziv (LZ) algorithm \cite{ZL78}. Once again, both results are tight in terms of the guesswork exponent, and are universal in the source distribution and the moment. 

A critical factor in guessing strategies is their implementation. Deterministic approaches require hard-coding long lists (exponentially large in the block length), and hence are memory consuming, while in a randomized approach, one needs to sample from a specific distribution, which might require computing an exponential sum. In this paper, we give two efficient algorithms to sample from the universal distribution we propose. The first algorithm is based on (a repeated) random walk on a growing tree, thus randomly and independently generating new LZ phrases, to be used as guesses. The second algorithm is based on feeding a (slightly modified) LZ decoder with purely random bits. Finally, the results and algorithms are extended to the case with side information. 

The rest of this paper is organized as follows. In \Cref{related work}, we review the current literature with references both to information--theoretic results and
to key findings regarding brute force attacks on passwords. In \Cref{notation and statement}, we formally define the problem and our objectives. \Cref{memoryless} describes the results for memoryless sources, while \Cref{finite state} describes the results for sources with memory. \Cref{conclusions} concludes the paper.

\section{Related Work}\label{related work}

The first information--theoretic study on guesswork was carried out by Massey \cite{Massey94}. Arikan \cite{Arikan96} showed, among other things, that the exponential rate of the number of guesses required for memoryless sources is given by the R\`enyi entropy of order $\frac{1}{2}$. Guesswork under a distortion constraint was studied by Arikan and Merhav \cite{AM98}, who also derived a guessing strategy for discrete memoryless sources (DMS's), which is
universally asymptotically optimal, both in the unknown memoryless source and the moment order of the guesswork being analyzed.

Guesswork for Markov processes was studied by Malone and Sullivan \cite{malone2004guesswork}, and extended to a large class of stationary measures by Pfister and Sullivan \cite{PfisterSullivan}. In \cite{5673955}, Hanawal and Sundaresan proposed a large deviations approach. They derived the guesswork exponent for sources satisfying a large deviations principle (LDP), and thereby generalized the results in \cite{Arikan96} and \cite{malone2004guesswork}. In \cite{christiansen2013guesswork}, again via large deviations, Christiansen \emph{et al.} proposed an approximation to the \emph{distribution} of the guesswork.
In \cite{sundaresan2007guessing}, Sundaresan considered guessing under source uncertainty. The redundancy as a function of the \emph{radius} of the family of possible distributions was defined and quantified in a few cases. For the special class of discrete memoryless sources, as already suggested in \cite{AM98}, this redundancy tends to zero as the length of the sequence grows without bound. 

In \cite{christiansen2015multi}, Christiansen \emph{et al.} considered a multi-user case, where an adversary (inquisitor) has to guess $U$ out of $V$ strings, chosen at random from some string-source $\mu^n$. Later, Beirami \emph{et al.} \cite{7282958} further defined the \emph{inscrutability} $S^n(U,V,\mu^n)$ of a string-source, the \emph{inscrutability rate} as the exponential rate of $S^n(U,V,\mu^n)$, and gave upper and lower bounds on this rate by identifying the appropriate string-source distributions. They also showed that ordering strings by their type-size in ascending order is a universal guessing strategy. Note, however, that both \cite{christiansen2015multi,7282958} considered a single attacker, with the ability to create a list of strings and guess one string after the other. 

Following Weinberger \emph{et al.} \cite{Weinberger_universal_ordering_92}, ordering strings by the size of their type-class before assigning them \emph{codewords in a fixed-to-variable source coding scheme} was also found useful by Kosut and Sankar in \cite{Kosut_universal_F2V_17} to minimize the third order term of the minimal number of bits required in lossless source coding (the first being the entropy, while the second is the dispersion). A geometric approach to guesswork was proposed by Beirami \emph{et al.} in \cite{Beirami15geometric}, showing that indeed the dominating type in guesswork (the position of a given string in the list) is the largest among all types whose elements are more likely than the given string. Here we show that a similar ordering is also beneficial for universal, \emph{decentralized guessing}, though the sequences are not ordered in practice, and merely assigned \emph{probabilities to be guessed} based on their type or LZ complexity.

Guesswork over a binary erasure channel was studied by Christiansen \emph{et al.} in \cite{christiansen2013guessing}. While the underlying sequence to be guessed was assumed i.i.d., the results therein apply to channels with memory as well (yet satisfying an LDP). Interestingly, it was shown that the guesswork exponent is higher than
the noiseless exponent times the average fraction of erased symbols, and one pays a non-negligible toll for the randomness in the erasures pattern.

In \cite{salamatian2017centralized}, Salamatian \emph{et al.} considered multi-agent guesswork with side information. The effect of \emph{synchronizing the side information} among the agents was discussed, and its affect on the exponent was quantified. Multi-agent guesswork was then also studied in \cite{SHBCM18}, this time devising a randomized guessing strategy, which can be used by asynchronous agents. The strategy in \cite{SHBCM18}, however, is hard to implement in practice, as it depends on both the source distribution and the moment of the guesswork considered, and requires computing an exponential sum.\footnote{An algorithm which produces a Markov chain, whose distribution converges to this sum was suggested in the context of randomized guessing tools in \cite{hanawal2010randomised}, yet only asymptotically and only for fixed, known distributions.}

Note that besides the standard application of guessing a password for a certain service, while knowing a password of the same user to another service, guessing with side information may also be applicable when breaking lists of hashed \emph{honeywords} \cite{juels2013honeywords,wang2018security}. In this scenario, an attacker is faced with a list of hashed sweatwords, where one is the hash of the true password while the rest are hashes of decoy honeywords, created \emph{with strong correlation} to the real password. If one is broken, using it as side information can significantly reduce the time required to break the others.
Furthermore, guessing with side information is also related the problem of \emph{guessing using hints} \cite{bracher2017guessing}. In this scenario, a legitimate decoder should be able to guess a password (alternatively, a \emph{task} to be carried out) using several hints, in the sense of having a low expected conditional guesswork, yet an eavesdropper knowing only a subset of the hints, should need a large number of guesses. In that case, the expected conditional guesswork generalizes secret sharing schemes by quantifying the amount of work Bob and Eve have to do. 

From a more practical viewpoint, trying to create passwords based on real data, Weir \emph{et al.} \cite{5207658} suggested a context-free grammar to create passwords at a descending order of probabilities, where the grammar rules as well as the probabilities of the generalized letters (sequences of English letters, sequences of digits or sequences of special characters) were learned based on a given training set. In \cite{dell2010password}, Dell'Amico \emph{et al.} evaluated experimentally the probability of guessing passwords using dictionary-based, grammar-free and Markov chain strategies, using existing data sets of passwords for validation. Not only was it clear that complex guessing strategies, which take into account the memory, perform better, but moreover, the authors stress out the \emph{need to fine-tune memory parameters} (e.g., the length of sub-strings tested), strengthening the necessity for a universal, parameter-free guessing strategy. In \cite{bonneau2012science} Bonneau also implicitly mentions the problems coping with passwords from an unknown distribution, or an unknown mixture of several known distributions.
\section{Notation Conventions, Problem Statement and Objectives}\label{notation and statement}
\subsection{Notation Conventions}
Throughout the paper, random variables will be denoted by capital
letters, specific values they may take will be denoted by the
corresponding lower case letters, and their alphabets
will be denoted by calligraphic letters. Random
vectors and their realizations will be denoted,
respectively, by capital letters and the corresponding lower case letters,
both in the bold face font. Their alphabets will be superscripted by their
dimensions. For example, the random vector $\bX=(X_1,\ldots,X_n)$, ($n$ --
positive integer) may take a specific vector value $\bx=(x_1,\ldots,x_n)$
in $\calX^n$, the $n$--th order Cartesian power of $\calX$, which is
the alphabet of each component of this vector.
Sources and channels will be denoted by the letter $P$ or $Q$ with or without some subscripts.
When there is no room for ambiguity, these
subscripts will be omitted.
The expectation
operator will be
denoted by
$\bE\{\cdot\}$.
The entropy of a generic distribution $Q$ on $\calX$ will be denoted by
$H_Q(X)$ where $X$ designates a random variable drawn by $Q$. For two
positive sequences $a_n$ and $b_n$, the notation $a_n\exe b_n$ will
stand for equality in the exponential scale, that is,
$\lim_{n\to\infty}\frac{1}{n}\log \frac{a_n}{b_n}=0$. Similarly,
$a_n\lexe b_n$ means that
$\limsup_{n\to\infty}\frac{1}{n}\log \frac{a_n}{b_n}\le 0$, and so on.
When both sequences depend on a vector, $\bx\in\calX^n$, namely, 
$a_n=a_n(\bx)$ and $b_n=b_n(\bx)$,
the notation $a_n(\bx)\exe b_n(\bx)$ means that the asymptotic convergence is
uniform, namely, 
$\lim_{n\to\infty}\max_{\bx\in\calX^n}|\frac{1}{n}\log
\frac{a_n(\bx)}{b_n(\bx)}|=0$. Likewise, $a_n(\bx)\lexe b_n(\bx)$
means $\limsup_{n\to\infty}\max_{\bx\in\calX^n}\frac{1}{n}\log
\frac{a_n(\bx)}{b_n(\bx)}\le 0$, and so on.

The empirical distribution of a sequence $\bx\in\calX^n$, which will be
denoted by $\hat{P}_{\bx}$, is the vector of relative frequencies
$\hat{P}_{\bx}(x)$
of each symbol $x\in\calX$ in $\bx$.
The type class of $\bx\in\calX^n$, denoted $\calT(\bx)$, is the set of all
vectors $\bx'$
with $\hat{P}_{\bx'}=\hat{P}_{\bx}$.
Information measures associated with empirical distributions
will be denoted with `hats' and will be subscripted by the sequences from
which they are induced. For example, the entropy associated with
$\hat{P}_{\bx}$, which is the empirical entropy of $\bx$, will be denoted by
$\hat{H}_{\bx}(X)$. 
Similar conventions will apply to the joint empirical
distribution, the joint type class, the conditional empirical distributions
and the conditional type classes associated with pairs of
sequences of length $n$.
Accordingly, $\hP_{\bx\by}$ would be the joint empirical
distribution of $(\bx,\by)=\{(x_i,y_i)\}_{i=1}^n$,
$\calT(\bx,\by)$ or $\calT(\hP_{\bx\by})$ will denote
the joint type class of $(\bx,\by)$, $\calT(\bx|\by)$ will stand for
the conditional type class of $\bx$ given $\by$,
$\hH_{\bx\by}(X|Y)$ will be the empirical conditional entropy,
and so on.

In Section IV, the broader notion of a type class, which
applies beyond the memoryless case, will be adopted: the type class
of $\bx$ w.r.t.\ a given class of sources $\calP$, will be defined as
\begin{equation}
\calT(\bx)=\bigcap_{P\in\calP}\{\bx^\prime:~P(\bx^\prime)=P(\bx)\}.
\end{equation}
Obviously, the various type classes, $\{\calT(\bx)\}_{\bx\in\calX^n}$, are
equivalence classes, and therefore, form a partition of $\calX^n$. Of course,
when $\calP$ is the class of memoryless sources over $\calX$, this definition
of $\calT(\bx)$ is equivalent to
the earlier one, provided in the previous paragraph.
\subsection{Problem Statement and Objectives}
In this paper, we focus on the guessing problem that is defined as follows.
Alice selects a secret random $n$--vector $\bX$,
drawn from a finite alphabet source $P$. Bob, which is unaware of the
realization of $\bX$, submits a sequence of guesses in the form of yes/no queries:
``Is $\bX=\bx_1$?''
``Is $\bX=\bx_2$?'', and so on, until receiving an affirmative answer.
A {\it guessing list}, $\calG_n$, is an ordered list of all members of $\calX^n$,
that is, $\calG=\{\bx_1,\bx_2,\ldots,\bx_M\}$, $M=|\calX|^n$, and it is associated
with a {\it guessing function}, $G(\bx)$, which is the function that maps $\calX^n$ onto
$\{1,2,\ldots,M\}$ by assigning to each $\bx\in\calX^n$ the integer $k$ for
which $\bx_k=\bx$, namely, the $k$--th element of $\calG_n$. In other words, $G(\bx)$ is the
number of guesses required until success, using $\calG_n$, when $\bX=\bx$.

The guessing problem is about devising a guessing list $\calG_n$ that minimizes
a certain moment of $G(\bX)$, namely, $\bE\{G^\rho(\bX)\}$, where $\rho > 0$ is
a given positive real (not necessarily a natural number). Clearly, when the
source $P$ is known and $\rho$ is arbitrary, 
the optimal guessing list orders the members of $\calX^n$
in the order of non--increasing probabilities. When $P$ is unknown, but known
to belong to a given parametric class $\calP$, like the class of memoryless sources,
or the class of finite--state sources with a given number of states, we are
interested in a {\it universal guessing list}, which is asymptotically optimal
in the sense of minimizing the {\it guessing exponent}, namely, achieving
\begin{equation}
E(\rho)=\limsup_{n\to\infty}\min_{\calG_n}\frac{\log \bE\{G^\rho(\bX)\}}{n},
\end{equation}
uniformly for all sources in $\calP$ and all positive real values of $\rho$.

Motivated by applications of distributed, asynchronous guessing by several
agents (see Introduction), we will also be interested in {\it randomized
guessing} schemes, which have the advantages of: (i) relaxing the need to consume large
volumes of memory (compared to deterministic guessing which needs the storage
of the guessing list $\calG_n$) and (ii) dropping the need for synchronization among
the various guessing agents (see \cite{SHBCM18}).
In randomized guessing, the guesser sequentially submits 
a sequence of random guesses, each one distributed independently according to a certain
probability distribution $\tilde{P}(\bx)$. 
We would like the distribution $\tilde{P}$ to be universally
asymptotically optimal in the sense of achieving (on the average) 
the optimal guessing exponent, while being
independent of the unknown source $P$ and independent of $\rho$.
Another desirable feature of the random guessing distribution $\tilde{P}$ 
is that it would be easy to implement
in practice. This is especially important when $n$ is large, as it is not
trivial to implement a general distribution over $\calX^n$ in the absence of
any structure to this distribution.

We begin our discussion from the case where the class of sources, $\calP$, is
the class of memoryless sources over a finite alphabet $\calX$ of
size $\alpha$. In this case, some of
the results we will mention are already known, but it will be helpful, as a
preparatory step, before we address the more interesting and challenging case,
where $\calP$ is the class of all non--unifilar, finite--state sources,
a.k.a.\ hidden
Markov sources (over the same finite alphabet $\calX$), 
where even the number of states is unknown to the guesser,
let alone, the parameters of the source for a given number of states.
In both cases, we also extend the study to the case where the guesser is
equipped with side information (SI) $\bY$, correlated to the vector $\bX$ to
be guessed.
\section{Guessing for Memoryless Sources}\label{memoryless}
\subsection{Background}
Following \cite{Massey94}, Arikan \cite{Arikan96} has established some
important bounds associated with guessing moments with relation the R\'enyi
entropy, with and without side information, where the main application he had
in mind was sequential decoding. Some of Arikan's results set the
stage for guessing $n$--vectors emitted from memoryless sources.
Some of these results were later extended to the 
case of lossy guessing\footnote{Here, by ``lossy guessing'' we mean that a
guess is considered successful if its distance (in the sense of a given
distortion function) from the underlying source vector, does not exceed a
given distortion level.}
\cite{AM98} with a certain emphasis
on universality issues. In particular, narrowing down the main result of
\cite{AM98} to the case of lossless guessing considered here, 
it was shown that the best achievable guessing exponent, $E(\rho)$,
is given by the following single--letter expression for a given memoryless
source $P$:
\begin{equation}
\label{gexp}
E(\rho)=\max_Q[\rho H_Q(X)-D(Q\|P)]=\rho H^{1/(1+\rho)}(X),
\end{equation}
where $Q$ is an auxiliary distribution over $\calX$ to be optimized, and
$H^\alpha(X)$ designates the R\'enyi entropy of order $\alpha$,
\begin{equation}
H^\alpha(X)=\frac{1}{1-\alpha}\ln\left[\sum_{x\in\calX}P^\alpha(x)\right],
\end{equation}
which is asymptotically achieved using a universal deterministic 
guessing list, $\calG_n$, that orders the
members of $\calX^n$ according to a non--decreasing order of their empirical
entropies, namely,
\begin{equation}
\hH_{\bx_1}(X)\le
\hH_{\bx_2}(X)\le\ldots\le
\hH_{\bx_M}(X).
\end{equation}
In the presence of correlated side information $\bY$, generated from $\bX$ by a
discrete memoryless channel (DMC), the above findings continue to hold, with
the modifications
that: (i) $H_Q(X)$ is replaced by $H_Q(X|Y)$, (ii) $D(Q\|P)$ is understood to be the
divergence between the two joint distributions of the pair $(X,Y)$
(which in turn implies that $H^{1/(1+\rho)}(X)$ is replaced by the 
corresponding conditional R\'enyi entropy of $X$ given $Y$), and (iii)
$\hH_{\bx_k}(X)$ is replaced by
$\hH_{\bx_k\by}(X|Y)$, $k=1,2,\ldots,M$.
\subsection{Randomized Guessing and its Efficient Implementation}
For universal randomized guessing, we consider the following guessing distribution
\begin{equation}
\label{univmemoryless}
\tilde{P}(\bx)=\frac{2^{-n\hH_{\bx}(X)}}{\sum_{\bx^\prime}
2^{-n\hH_{\bx^\prime}(X)}}.
\end{equation}
We then have the following result:
\begin{theorem}
Randomized guessing according to eq.\ (\ref{univmemoryless}) achieves the
optimal guessing exponent (\ref{gexp}).
\end{theorem}

\noindent
{\it Proof.} We begin from the following lemma, whose proof is deferred 
to the appendix.
\begin{lemma}
\label{geometricsum}
For given $a \ge 0$ and $\rho>0$,
\begin{equation}
\sum_{k=1}^\infty k^\rho (1-e^{-na})^{k-1}\lexe e^{(1+\rho)na}.
\end{equation}
\end{lemma}

Denoting by $\calP_n$ the set of probability distributions over $\calX$
with rational letter probabilities of denominator $n$ (empirical
distributions), we
observe that since
\begin{eqnarray}
1&\le&\sum_{\bx}2^{-n\hH_{\bx}(X)}\nonumber\\
&=&\sum_{\bx}\max_{Q\in\calP}Q(\bx)\nonumber\\
&=&\sum_{\bx}\max_{Q\in\calP_n}Q(\bx)\nonumber\\
&\le&\sum_{\bx}\sum_{Q\in\calP_n}Q(\bx)\nonumber\\
&=&\sum_{Q\in\calP_n}\sum_{\bx}Q(\bx)\nonumber\\
&=&|\calP_n|\nonumber\\
&\le&(n+1)^{|\calX|-1},
\end{eqnarray}
it follows that
\begin{equation}
\tilde{P}(\bx)\exe 2^{-n\hH_{\bx}(X)}.
\end{equation}
Given that $\bX=\bx$, the $\rho$--th moment of the number of guesses under $\tilde{P}$ is
given by
\begin{eqnarray}
\label{conditionalmoment}
& &\sum_{k=1}^\infty k^\rho [1-\tilde{P}(\bx)]^{k-1}\tilde{P}(\bx)\nonumber\\
&=&\tilde{P}(\bx)\sum_{k=1}^\infty k^\rho [1-\tilde{P}(\bx)]^{k-1}\nonumber\\
&\exe&2^{-n\hH_{\bx}(X)}\sum_{k=1}^\infty k^\rho [1-2^{-n\hH_{\bx}(X)}]^{k-1}\nonumber\\
&\lexe&2^{-n\hH_{\bx}(X)}2^{n(1+\rho)\hH_{\bx}(X)}\nonumber\\
&=&2^{n\rho\hH_{\bx}(X)},
\end{eqnarray}
where in the inequality, we have used Lemma \ref{geometricsum} with
the assignment $a=\hH_{\bx}(X)\ln 2$.
Taking the expectation of $2^{n\rho\hH_{\bx}(X)}$ w.r.t.\ $P(\bx)$, 
using the method of types \cite{CK11},
one easily obtains (see also \cite{AM98}) 
the exponential order of $2^{nE(\rho)}$, with $E(\rho)$ as
defined in (\ref{gexp}). This completes the proof of Theorem 1. $\Box$\\

\noindent
{\it Remark:}
It is easy to see that the random guessing scheme has an additional important feature: not only the expectation of $G^\rho(\bx)$ (w.r.t.\ the randomness of the guesses) has the optimal exponential order of $2^{n\rho\hH_{\bx}(X)}$ for each and every $\bx$, but moreover, the probability that $G(\bx)$ would exceed
$2^{n[\hH_{\bx}(X)+\epsilon]}$ decays double--exponentially rapidly for every $\epsilon > 0$. This follows from the following
simple chain of inequalities:
\begin{eqnarray}
\mbox{Pr}\left\{G(\bx)\ge 2^{n[\hH_{\bx}(X)+\epsilon]}\right\}&\exe&\left[1-2^{-n\hH_{\bx}(X)}\right]^{2^{n[\hH_{\bx}(X)+\epsilon]}}\nonumber\\
&=&\exp\left\{2^{n[\hH_{\bx}(X)+\epsilon]}\ln\left[1-2^{-n\hH_{\bx}(X)}\right]\right\}\nonumber\\
&\le&\exp\left\{-2^{n[\hH_{\bx}(X)+\epsilon]}\cdot2^{-n\hH_{\bx}(X)}\right\}\nonumber\\
&=&\exp\left\{-2^{n\epsilon}\right\}.
\end{eqnarray}
A similar comment will apply also to the random guessing scheme of Section 5.\\

The random guessing distribution (\ref{univmemoryless})
is asymptotically equivalent (in the exponential scale)
to a class of mixtures of all memoryless sources
over $\calX$, having the form
\begin{equation}
M(\bx)=\int_{\calS}\mu(Q)\cdot Q(\bx)\mbox{d}Q
\end{equation}
where $\mu(\cdot)$ is a density defined on the simplex of all distributions on
$\calX$, and where it is assumed that $\mu(\cdot)$ is bounded away from zero
and from infinity, and that it is independent of $n$.
As mentioned in \cite{KT81}, one of the popular
choices of $\mu(\cdot)$ is the Dirichlet distribution, parametrized by
$\lambda > 0$,
\begin{equation}
\mu(Q)=\frac{\Gamma(\lambda|\calX|)}{\Gamma^{|\calX|}(\lambda)}
\cdot\left[\prod_{x\in\calX}
Q(x)\right]^{\lambda-1},
\end{equation}
where
\begin{equation}
\Gamma(s)\dfn \int_0^\infty x^{s-1}e^{-x}\mbox{d}x,
\end{equation}
and we remind that for a positive integer $n$,
\begin{eqnarray}
\Gamma(n)&=&(n-1)!\\
\Gamma\left(\frac{1}{2}\right)&=&\sqrt{\pi}\\
\Gamma\left(n+\frac{1}{2}\right)&=&\sqrt{\pi}\cdot\frac{1}{2}\cdot\frac{3}{2}\cdot
\cdot\cdot\left(n-\frac{1}{2}\right),~~~~n\ge 1.
\end{eqnarray}
For example, with the choice $\lambda=1/2$, the mixture becomes
\begin{equation}
M(\bx)=\Gamma\left(\frac{|\calX|}{2}\right)\cdot
\frac{\prod_{x\in\calX}\Gamma(n\hat{P}_{\bx}(x)+1/2)}
{\pi^{|\calX|/2}\Gamma(n+|\calX|/2)}.
\end{equation}
This mixture distribution can be implemented sequentially, as
\begin{equation}
M(\bx)=\prod_{t=0}^{n-1}M(x_{t+1}|x^t),
\end{equation}
where
\begin{equation}
M(x_{t+1}|x^t)=\frac{M(x^{t+1})}{M(x^t)}=\frac{t\hat{P}_{x^t}(x_{t+1})
+1/2}{t+|\calX|/2},
\end{equation}
where $\hat{P}_{x^t}(x)$ is the relative frequency of $x\in\calX$ in
$x^t=(x_1,\ldots,x_t)$. So the sequential implementation is rather simple:
draw the first symbol, $X_1$, according to the uniform distribution. Then, for
$t=1,2,\ldots,n-1$, draw the next symbol, $X_{t+1}$, according to the last
equation, taking into account the relative frequencies of the various letters
drawn so far.
\subsection{Side Information}
All the above findings extend straightforwardly to the case of a guesser that
is equipped with SI $\bY$, correlated to the random vector $\bX$ to be guessed,
where it is assumed that $(\bX,\bY)$ is a sequence of $n$ independent copies
of a pair of random variables $(X,Y)$ jointly distributed according to
$P_{XY}$. 

The only modification required is that the 
universal randomized guessing distribution will now by
proportional (and exponentially equivalent) to $2^{-n\hH_{\bx\by}(X|Y)}$
instead of $2^{-n\hH_{\bx}(X)}$, and in the sequential implementation, the
mixture and hence also the relative frequency counts will be applied to each SI letter
$y\in\calY$ separately. Consequently, the conditional distribution 
$M(x_{t+1}|x^t)$ above would be replaced by
\begin{equation}
M(x_{t+1}|x^t,y^t)=\frac{t\hat{P}_{x^ty^t}(x_{t+1},y_{t+1})
+1/2}{t\hat{P}_{y^t}(y_{t+1})+|\calX|/2}.
\end{equation}
\section{Guessing for Finite--State Sources}\label{finite state}
We now extend the scope to a much more general class of sources -- the class
of non--unifilar finite--state sources, namely, hidden Markov sources
\cite{EM02}. Specifically, we assume that $\bX$ is drawn by a distribution $P$
given by
\begin{equation}
\label{hmm}
P(\bx)=\sum_{\bz}\prod_{i=1}^nP(x_i,z_{i+1}|z_i),
\end{equation}
where $\{x_i\}$ is the source sequence as before, whose elements take 
on values in a finite alphabet $\calX$ of size $\alpha$, and 
where $\{z_i\}$ is the underlying state sequence, whose elements take on
values in a finite set of states, $\calZ$ of size $s$, and
where the initial state, $z_1$, is assumed to be a fixed member of $\calZ$.
The parameter set $\{P(x,z^\prime|z),x\in\calX,~z,z^\prime\in\calZ\}$
is unknown the guesser. In fact, even the number of states, $s$, is not known, and we
seek a universal guessing strategy.
\subsection{Converse Theorem}
Let us parse $\bx$ into $c=c(\bx)$ distinct phrases,
by using,
for example, the incremental parsing procedure\footnote{The
incremental parsing procedure is a sequential procedure of parsing a sequence,
such that each new parsed phrase is the shortest string that has not been
obtained before as a phrase.}
of the Lempel--Ziv (LZ)
algorithm \cite{ZL78} (see also \cite[Subsection 13.4.2]{CT06}).
The following is a converse theorem concerning the best achievable guessing
performance.

\begin{theorem}
\label{converse}
Given a finite--state source (\ref{hmm}), any guessing function satisfies the
following inequality:
\begin{equation}
\bE\{G^\rho(\bX)\}\ge 2^{-n\Delta_n}\bE\left[\exp_2\{\rho
c(\bX)\log c(\bX)\}\right],
\end{equation}
where $\Delta_n$ is a function of $s$, $\alpha$ and $n$, that tends to zero as
$n\to\infty$ for fixed $s$ and $\alpha$.
\end{theorem}

\noindent
{\it Proof.}
Without essential loss of generality, let $\ell$ divide $n$ and consider the segmentation of
$\bx=(x_1,\ldots,x_n)$ into $n/\ell$ non--overlapping sub--blocks,
$\bx_i=(x_{i\ell+1},x_{i\ell+2},\ldots,x_{(i+1)\ell})$,
$i=0,1,\ldots,n/\ell-1$.
Let $\bz^\ell =(z_1,z_{\ell+1},z_{2\ell+1},\ldots,z_{n+1})$ be the
(diluted) state sequence pertaining to the boundaries between neighboring
sub--blocks. Then,
\begin{equation}
P(\bx,\bz^\ell)=\prod_{i=0}^{n/\ell-1}P(\bx_i,z_{(i+1)\ell+1}|z_{i\ell+1}).
\end{equation}
For a given $\bz^\ell$, let $\calT(\bx|\bz^\ell)$ be the set of all sequences
$\{\bx^\prime\}$ that are obtained by permuting different sub--blocks that both
begin at the same state and end at the same state. Owing to the product form of
$P(\bx,\bz^\ell)$, it is clear that $P(\bx^\prime,\bz^\ell)=
P(\bx,\bz^\ell)$ whenever $\bx^\prime\in\calT(\bx|\bz^\ell)$. It was shown in
\cite[Eq.\ (47) and Appendix A]{me91} that
\begin{equation}
|\calT(\bx|\bz^\ell)|\ge \exp_2\{c(\bx)\log c(\bx)-n\delta(n,\ell)\},
\end{equation}
independently of $\bz^\ell$, where $\delta(n,\ell)$ tends to $C/\ell$ ($C> 0$
-- constant) as
$n\to\infty$ for fixed $\ell$. Furthermore, by choosing
$\ell=\ell_n=\sqrt{\log n}$, we have that
$\delta(n,\ell_n)=O(1/\sqrt{\log n})$. We then have the following chain of
inequalities:
\begin{eqnarray}
\bE\{G^\rho(\bX)\}&=&\sum_{\bz^{\ell_n}}\sum_{\bx}P(\bx,\bz^{\ell_n})G^\rho(\bx)\nonumber\\
&=&\sum_{\bz^{\ell_n}}\sum_{\{T(\bx|\bz^{\ell_n})\}}\sum_{\bx^\prime\in\calT(\bx|\bz^{\ell_n})}
P(\bx,\bz^{\ell_n})G^\rho(\bx)\nonumber\\
&=&\sum_{\bz^{\ell_n}}\sum_{\{T(\bx|\bz^{\ell_n})\}}P(\bx,\bz^{\ell_n})
\sum_{\bx^\prime\in\calT(\bx|\bz^{\ell_n})}
G^\rho(\bx)\nonumber\\
&=&\sum_{\bz^{\ell_n}}\sum_{\{T(\bx|\bz^{\ell_n})\}}
P(\bx,\bz^{\ell_n})\cdot|\calT(\bx|\bz^{\ell_n})|\cdot
\sum_{\bx^\prime\in\calT(\bx|\bz^{\ell_n})}
\frac{G^\rho(\bx)}{|\calT(\bx|\bz^{\ell_n})|}\nonumber\\
&\ge&\sum_{\bz^{\ell_n}}\sum_{\{T(\bx|\bz^{\ell_n})\}}P(\bx,\bz^{\ell_n})\cdot
|\calT(\bx|\bz^{\ell_n})|\cdot
\frac{|\calT(\bx|\bz^{\ell_n})|^\rho}{1+\rho}\nonumber\\
&\ge&\frac{1}{1+\rho}\sum_{\bz^{\ell_n}}\sum_{\{T(\bx|\bz^{\ell_n})\}}
P(\bx,\bz^{\ell_n})\cdot|\calT(\bx|\bz^{\ell_n})|\cdot
\exp_2\{\rho[c(\bx)\log c(\bx)-n\delta(n,\ell_n)]\}\nonumber\\
&=&\frac{2^{-n\delta(n,\ell_n)}}{1+\rho}\bE\left[\exp_2\{\rho\cdot
c(\bX)\log c(\bX)\}\right]\nonumber\\
&\dfn&2^{-n\Delta_n}\bE\left[\exp_2\{\rho\cdot
c(\bX)\log c(\bX)\}\right],
\end{eqnarray}
where the first inequality follows from the following genie--aided argument:
The inner--most summation in the fourth line of the above chain can be viewed
as the guessing moment of a guesser that is {\it informed} that $\bX$ falls within a
given $\calT(\bx|\bz^{\ell_n})$. Since the distribution within
$\calT(\bx|\bz^{\ell_n})$ is uniform, no matter what guessing strategy may be,
\begin{eqnarray}
\sum_{\bx^\prime\in\calT(\bx|\bz^{\ell_n})}
\frac{G^\rho(\bx)}{|\calT(\bx|\bz^{\ell_n})|}&\ge&\frac{1}{|\calT(\bx|\bz^{\ell_n})|}
\sum_{k=1}^{|\calT(\bx|\bz^{\ell_n})|}k^\rho\nonumber\\
&\ge&|\calT(\bx|\bz^{\ell_n})|^\rho\cdot\frac{1}{|\calT(\bx|\bz^{\ell_n})|}
\sum_{k=1}^{|\calT(\bx|\bz^{\ell_n})|}
\left(\frac{k}{|\calT(\bx|\bz^{\ell_n})|}\right)^\rho\nonumber\\
&\ge&|\calT(\bx|\bz^{\ell_n})|^\rho\cdot\int_0^1 u^\rho\mbox{d}u\nonumber\\
&=&\frac{|\calT(\bx|\bz^{\ell_n})|^\rho}{1+\rho}.
\end{eqnarray}
This completes the proof of Theorem \ref{converse}. $\Box$
\subsection{Direct Theorem}
We now present a matching direct theorem, which asymptotically achieves the
converse bound in the exponential scale.
\begin{theorem}
\label{direct}
Given a finite--state source (\ref{hmm}), 
there exists a universal guessing list that satisfies the
following inequality:
\begin{equation}
\bE\{G^\rho(\bX)\}\le 2^{n\Delta_n^\prime}\bE\left[\exp_2\{\rho
c(\bX)\log c(\bX)\}\right],
\end{equation}
where $\Delta_n^\prime$ is a function of $s$, $\alpha$ and $n$, that tends to zero as
$n\to\infty$ for fixed $s$ and $\alpha$.
\end{theorem}

\noindent
{\it Proof.}
The proposed deterministic guessing list orders all members of $\calX^n$ in
non--decreasing order of their Lempel--Ziv code--lengths
\cite[Theorem 2]{ZL78}. Denoting the LZ code--length of $\bx$ by $LZ(\bx)$,
we then have
\begin{eqnarray}
G(\bx)&\le&|\{\bx^\prime:~LZ(\bx^\prime)\le
LZ(\bx)\}|\nonumber\\
&=&\sum_{i=1}^{LZ(\bx)}|\{\bx^\prime:~LZ(\bx^\prime)=i\}|\nonumber\\
&\le&\sum_{i=1}^{LZ(\bx)} 2^i\nonumber\\
&<& 2^{LZ(\bx)+1}\nonumber\\
&\le& \exp_2\{[c(\bx)+1]\log(2\alpha[c(\bx)+1])+1\}\nonumber\\
&\exe&\exp_2\{c(\bx)\log c(\bx)\},
\end{eqnarray}
where the inequality $|\{\bx^\prime:~LZ(\bx^\prime)=i\}|\le 2^i$ is due to the
fact that the LZ code is uniquely decipherable (UD) and the last inequality is from
Theorem 2 of \cite{ZL78}.
By raising this inequality to the power of $\rho$ and taking the expectation
of both sides, Theorem \ref{direct} is readily proved. 

An alternative, randomized guessing strategy pertains to independent random
guesses according to the following universal distribution,
\begin{equation}
\label{Plz}
\tilde{P}(\bx)=\frac{2^{-LZ(\bx)}}{\sum_{\bx^\prime}2^{-LZ(\bx^\prime)}}.
\end{equation}
Since the LZ code is UD, it satisfies the Kraft inequality, and so, the
denominator cannot be larger than 1, which means that
\begin{equation}
\label{lb}
\tilde{P}(\bx)\ge 2^{-LZ(\bx)}\ge
\exp_2\{-[c(\bx)+1]\log(2\alpha[c(\bx)+1])\}.
\end{equation}
Similarly as in (\ref{conditionalmoment}), applying Lemma \ref{geometricsum} 
to (\ref{Plz}) (or to (\ref{lb})), this time with
$a=\frac{\ln 2}{n}[c(\bx)+1]\log(2\alpha[c(\bx)+1])$, we obtain that the $\rho$--th
moment of $G(\bX)$ given that $\bX=\bx$ is upper bounded by an expression of the exponential of
$\exp_2\{\rho [c(\bx)+1]\log(2\alpha[c(\bx)+1])\}\exe \exp_2\{\rho c(\bx)\log
c(\bx)\}$, and then
upon taking the expectation w.r.t.\ 
the randomness of $\bX$, we readily obtain the achievability result once again. $\Box$
\subsection{Algorithms for Sampling From the Universal Guessing
Distribution}
Similarly as in Section 4, we are interested in efficient algorithms for
sampling from the universal distribution, (\ref{Plz}). In fact, it is enough
to have an efficient implementation of an algorithm that efficiently samples
from a distribution $\tilde{P}$ that satisfies $\tilde{P}(\bx)\gexe
2^{-c(\bx)\log c(\bx)}$.
We propose two different algorithms, the first is inspired by the predictive
point of view associated with LZ parsing \cite{Feder91}, \cite{FMG92}, and the
second one is based on the simple idea of feeding the LZ decoder with purely
random bits. The latter algorithm turns out to lend itself more easily to
generalization for the case of
guessing in the presence of SI. Both algorithms are described in terms of
walks on a growing tree, but the difference is that in the first algorithm, the tree
is constructed in the domain of the source sequences, whereas in the second
algorithm, the tree is in the domain of the compressed bit-stream.\\

\noindent
{\bf First algorithm.}
As said, the idea is in the spirit of the predictive probability assignment
mechanism proposed in
\cite{Feder91} and
\cite[Sect.\ V]{FMG92}, but here, instead of using the
incremental parsing
mechanism for prediction, we use it for random selection. 

As mentioned before, the algorithm is described as a process generated by a repeated walk on
a growing tree, beginning, each time, from the root and ending at 
one of the leaves. Consider a tree which is initially
composed of a root connected to
$\alpha$ leaves, each one corresponding to one alphabet
letter, $x\in\calX$. We always assign to each leaf 
a weight of 1 and to each internal node -- the sum of weights
of its immediate off-springs, and so, the initial weight of the root is $\alpha$. 
We begin by drawing the first symbol, $X_1$, such that the probability of
$X_1=x$ is given by the weight of $x$ (which is 1) divided by the weight of
the current node, which is the root (i.e., a weight of $\alpha$, as said). In
other words, we randomly select $X_1$ 
according to the uniform
distribution over $\calX$. The leaf corresponding to the outcome 
of $X_1$, call it $x_1$, will now become an internal node by adding to the
tree its $\alpha$ off-springs, thus growing the tree to have $2\alpha-1$ leaves.
Each one of the leaves of the extended tree has now weight 1, and then, the weight of the
their common ancestor (formerly, the leaf of $x_1$), becomes the sum of their
weights, namely, $\alpha$, and similarly, the weights of all ancestors of
$x_1$, all the way up to the root, 
are now sequentially updated to become the sum of the weights of their
immediate off-springs. 

We now start again
from the root of the
tree to randomly draw the next symbol, $X_2$, 
such that the probability that $X_2=x$, is given by the weight of the
node $x$ divided by the weight of the current node, which is again the root, 
and then we move from
the root to its corresponding off-spring pertaining to $X_2$, that was just randomly drawn.
If we have reached a leaf,
then again this leaf gives birth to $\alpha$ new off-springs, each assigned
with weight 1, then all
corresponding weights are updated as described before, and finally, 
we move back to the root, etc. 
If we are still in an internal node, then again, we draw the next symbol
according to the ratio between the weight of the node pertaining to 
the next symbol and the weight of the current node, and so on. 
The process continues until $n$
symbols, $X_1,X_2,\ldots,X_n$ have been generated.

Note that
every time we restart from the root and move along the tree until we reach a
leaf, we generate a new LZ phrase that has not been obtained before. Let $c(\bx)$
be the number of phrases generated. Along each path
from the root to a leaf, we implement a telescopic product of conditional
probabilities, where the numerator pertaining to the last conditional probability is the weight of the leaf,
which is 1, and the denominator of the first probability is the total number of
leaves after $i$ rounds, which is $\alpha+i(\alpha-1)$ (because after
every birth of a new generation of leaves, the total number of leaves is
increased by $\alpha-1$). All other numerators and denominators of the conditional
probabilities along the path cancel each other telescopically. The result is that the induced probability distribution along the various leaves is uniform. Precisely, after $i$ phrases have been generated, the probability of each leaf is exactly
$1/[\alpha+i(\alpha-1)]$.
Therefore,
\begin{eqnarray}
P(\bx)&=&\prod_{i=0}^{c(\bx)-1}\frac{1}{\alpha+i(\alpha-1)}\nonumber\\
&\ge&\prod_{i=0}^{c(\bx)-1}\frac{1}{\alpha+[c(\bx)-1](\alpha-1)}\nonumber\\
&=&\frac{1}{\{\alpha+[c(\bx)-1](\alpha-1)\}^{c(\bx)}}\nonumber\\
&=&2^{-c(\bx)\log\{[c(\bx)-1](\alpha-1)+\alpha\}},
\end{eqnarray}
which is of the exponential order of $2^{-c(\bx)\log c(\bx)}$.\\

\noindent
{\bf Second algorithm.}
As said, the second
method for efficiently generating random guesses according to the LZ distribution 
is based on the simple idea of feeding purely random bits into the LZ decoder 
until a decoded sequence of length $n$ is obtained. 
To describe it, we refer to the coding scheme proposed in \cite[Theorem
2]{ZL78}, but with a slight modification.
Recall that according to this coding scheme, 
for the $i$--th parsed phrase,
$\bx_{n_{j-1}+1}^{n_j}$, one encodes two integers: the 
index $0\le \pi(j)\le j-1$ of the matching past 
phrase and the index the additional source symbol, 
$0\le I_A(x_{n_j})\le \alpha-1$. 
These two integers are mapped together bijectively into one integer,
$I(\bx_{n_{j-1}+1}^{n_j})=\pi(j)\cdot\alpha+I_A(x_{n_j})$, which takes on values in the set
$\{0,1,2,\ldots,j\alpha-1\}$, and so, according to \cite{ZL78}, it can be encoded
using $L_j=\lceil\log(j\alpha)\rceil$ bits. 
Here, instead, we will encode $I(\bx_{n_{j-1}+1}^{n_j})$ 
with a tiny modification that will make the encoding equivalent 
to a walk on a {\it complete binary tree}
\footnote{By ``complete binary tree'', we mean a binary tree 
where each node is either a leaf or has two off-springs. The reason for the
need of a complete binary tree is that for the algorithm to be valid, 
every possible sequence of randomly chosen bits must be a
legitimate compressed bit-stream so that it would be decodable by the LZ
decoder.}
from the root to a leaf. Considering the fact that (by definition of $L_j$),
$2^{L_j-1}< j\alpha\le 2^{L_j}$, we first construct a full binary 
tree with $2^{L_j-1}$ leaves at depth $L_j-1$, and then 
convert $j\alpha-2^{L_j-1}$ leaves to internal nodes by generating their off-springs. 
The resulting complete binary tree will then have exactly $j\alpha$ leaves, 
some of them at depth $L_j-1$ and some - at depth $L_j$. 
Each leaf of this tree will now correspond to one value of
$I(\bx_{n_{j-1}+1}^{n_j})$, and hence to a certain decoded phrase. 
Let $\hat{L}_j$ denote the length of the
codeword for $I(\bx_{n_{j-1}+1}^{n_j})$. Obviously, either $\hat{L}_j=L_j-1$ or $\hat{L}_j=L_j$.
Consider now what happens if we feed the decoder of 
this encoder by a sequence of purely random bits (generated by a binary
symmetric source): every leaf at depth $\hat{L}_j$ will be obtained with 
probability $2^{-\hat{L}_j}$, and since the tree is complete, 
these probabilities sum up to unity. The probability of obtaining $\bx$ at 
the decoder output is, therefore, equal to the probability of the sequence of 
bits pertaining to its compressed form, namely, 
\begin{eqnarray}
\tilde{P}(\bx)&=&\prod_{j=1}^{c(\bx)+1} 2^{-\hat{L}_j}\nonumber\\
&=&\exp_2\left\{-\sum_{j=1}^{c(\bx)+1}\hat{L}_j\right\}\nonumber\\
&\ge&\exp_2\left\{-\sum_{j=1}^{c(\bx)+1}L_j\right\}\nonumber\\
&\ge&\exp_2\left\{-\sum_{j=1}^{c(\bx)+1}\log(2j\alpha)\right\}\nonumber\\
&\ge& \exp_2\left\{-[c(\bx)+1]\log [2c(\bx)\alpha]\right\}\nonumber\\
&=&\exp_2\{-[c(\bx)+1]\log c(\bx)-[c(\bx)+1]\log(2\alpha)\},
\end{eqnarray}
which is again of the exponential order of $2^{-c(\bx)\log c(\bx)}$.
\subsection{Side Information}
As we have done at the end of Section 4, here too,
we describe how our results extend to the case where the
guesser is equipped with SI. The parts that extend straightforwardly will be described briefly, 
whereas the parts whose extension is non--trivial will be more detailed.

Consider the pair process $\{(X_t,Y_t)\}$, 
jointly distributed according to a hidden Markov model,
\begin{equation}
\label{joint distribution (x,y)}
P(\bx,\by)=\sum_{\bz}\prod_{t=1}^n P(x_t,y_t,z_{t+1}|z_t),
\end{equation}
where, as before, $z_t$ is the state at time $t$, 
taking on values in a finite set of states $\calZ$ of cardinality $s$. 

Here, our objective is to guess $\bx$ when $\by$ is available to the guesser
as SI. Most of
our earlier results extend quite easily to this case. Basically, the only
modification needed is to replace the LZ complexity of $\bx$ by the
conditional LZ complexity of $\bx$ given $\by$, which is defined as
in \cite{UK03} and \cite{Ziv85}. In particular,
consider the joint parsing of the sequence of pairs, 
$\{(x_1,y_1),(x_2,y_2),\ldots,(x_n,y_n)\}$, let $c(\bx,\by)$ denote 
the number of phrases, $c(\by)$ -- the number of distinct $\by$-phrases,
$\by(\ell)$ -- the $\ell$-th distinct $\by$-phrase, $1\le\ell\le c(\by)$, and finally, let
$c_{\ell}(\bx|\by)$ denote the number of times
$\by(\ell)$ appears as a phrase, or, equivalently, 
the number of distinct $\bx$-phrases that appear jointly with $\by(\ell)$, 
so that $\sum_{\ell=1}^{c(\by)}c_\ell(\bx|\by)=c(\bx,\by)$. Then, we define
\begin{equation}
\label{def. LZ(x|y)}
u(\bx|\by)=\sum_{\ell=1}^{c(\by)}c_\ell(\bx|\by)\log c_\ell(\bx|\by).
\end{equation}

For the converse theorem (lower bound), the proof is the same as the proof of
Theorem \ref{converse}, except that here, we need
a lower bound on the size of a 
``conditional type'' of $\bx$ given $\by$. This lower bound turns to be of the 
exponential order of  $2^{u(\bx|\by)}$, as can be seen in
\cite[Lemma 1]{me00}. Thus, the lower bound on the guessing moment is of the
exponential order of $\bE[\exp_2\{\rho u(\bX|\bY)\}]$.

For the direct theorem (upper bound), we can either create a 
deterministic guessing list by ordering the members of $\calX^n$ according to 
increasing order of their conditional LZ code--length function values,
$LZ(\bx|\by)\approx u(\bx|\by)$, \cite[p.\ 2617]{UK03},
\cite[page 460, proof of Lemma 2]{Ziv85},
or randomly draw guesses according to
\begin{equation}
\label{def. Plz(x|y)}
\tilde{P}(\bx|\by)=\frac{2^{-LZ(\bx|\by)}}{\sum_{\bx'}2^{-LZ(\bx'|\by)}}.
\end{equation}
Following Subsection 5.C, we wish to have an efficient algorithm for sampling
from the distribution (\ref{def. Plz(x|y)}), or, more generally, for implementing a
conditional distribution that satisfies $\tilde{P}(\bx|\by)\gexe
2^{-LZ(\bx|\by)}\exe 2^{-u(\bx|\by)}$.

While we have not been able to find an extension of the first algorithm of Subsection
5.C to the case of SI, the second algorithm therein turns out to lend itself 
fairly easily to such
an extension. Once again, generally speaking, the idea is to feed a sequence
of purely random bits as inputs
to the decoder pertaining to the conditional LZ decoder, equipped with $\by$ as
SI, and wait until exactly $n$ symbols, $x_1,\ldots,x_n$, 
have been obtained at the output of the
decoder. We need, however, a few slight modifications in conditional LZ code, 
in order to ensure that any
sequence of randomly drawn bits would be legitimate as the output of the
encoder, and hence be also decodable by the decoder. Once again, to this end,
we must use complete binary trees for the prefix codes for the various
components of the conditional LZ code.

As can be seen in \cite{UK03}, \cite{Ziv85}, the conditional LZ compression algorithm
sequentially encodes $\bx$ phrase by phrase, where the code for each phrase
consists of three parts:
\begin{enumerate}
\item A code for the length of the phrase, $L[\by(\ell)]$.
\item A code for the location of the matching $\bx$--phrase among all previous
phrases with the same $\by$--phrase.
\item A code for the index of the last symbol of the $\bx$--phrase among all
members of $\calX$.
\end{enumerate}
Parts 2 and 3 are similar to those of the ordinary LZ algorithm and they in
fact can even be
united, as described before, into a single code for both indices (although
this is not necessary). Part 1 requires a code for the integers, which can be
implemented by the Elias code, as described in \cite{UK03}. However, for the sake
of conceptual simplicity of describing the required complete binary tree,
consider the following alternative option. Define the following distribution
on the natural numbers,
\begin{equation}
Q(i)=\frac{6}{\pi^2i^2},~~~~i=1,2,3,\ldots
\end{equation}
and construct a prefix tree for the corresponding Shannon code, whose length function is given by
\begin{equation}
\label{Li}
\calL(i)=\lceil -\log Q(i)\rceil.
\end{equation}
Next prune the tree by eliminating all leaves that correspond to values of
$i=L[\by(\ell)]$
that cannot be obtained at the current phrase: the length $L[\by(\ell)]$
cannot be larger than the maximum possible phrase length and cannot correspond
to a string that has not been obtained as a $\by$--phrase
before.\footnote{This is doable since both the encoder and the decoder have
this information at the beginning of the current phrase.} Finally, shorten the
tree by eliminating branches that emanate from any node that has one
off-spring only. At the end of this process, we have a complete binary tree
where the resulting code length for
every possible value of $L[\by(\ell)]$ cannot be larger than its original
value (\ref{Li}). 

The probability of obtaining a given $\bx$ at the output
of the above--described conditional LZ decoder is equal to the probability of
randomly selecting the bit-stream that generates $\bx$ (in the presence of
$\by$ as SI), as the response to this bit-stream.
Thus,
\begin{eqnarray}
\tilde{P}(\bx|\by)&\ge&\prod_{\ell=1}^{c(\by)}\prod_{j=1}^{c_\ell(\bx|\by)}
\exp_2\{-\lceil \log(j\alpha)\rceil-\calL(L[\by(\ell)])\}\nonumber\\
&\ge&\exp_2\left\{-\sum_{\ell=1}^{c(\by)}\sum_{j=1}^{c_\ell(\bx|\by)}\left[
\log(2j\alpha)+2\log
L[\by(\ell)]+\log\frac{\pi^2}{6}+1\right]\right\}\nonumber\\
&\ge&\exp_2\left\{-\sum_{\ell=1}^{c(\by)}c_\ell(\bx|\by)\log[2\alpha
c_\ell(\bx|\by)]-2\sum_{\ell=1}^{c(\by)}c_\ell(\bx|\by)\log
L[\by(\ell)] -\right.\nonumber\\
& &\left. c(\bx,\by)\left[\log\frac{\pi^2}{6}+1\right]\right\}\nonumber\\
&\exe& \exp_2\{-u(\bx|\by)\},
\end{eqnarray}
where the last step follows from the observation \cite[p.\ 460]{Ziv85}
that
\begin{eqnarray}
\sum_{\ell=1}^{c(\by)}c_\ell(\bx|\by)\log
L[\by(\ell)]
&=&c(\bx,\by)\sum_{\ell=1}^{c(\by)}\frac{c_\ell(\bx|\by)}{c(\bx,\by)}\log
L[\by(\ell)]\nonumber\\
&\le&c(\bx,\by)\log\left[\frac{\sum_{\ell=1}^{c(\by)}
c_\ell(\bx|\by)L[\by(\ell)]}{c(\bx,\by)}\right]\nonumber\\
&=&c(\bx,\by)\log \frac{n}{c(\bx,\by)},
\end{eqnarray}
and the fact that $c(\bx,\by)$ cannot be larger than $O(n/\log n)$
\cite{ZL78}.
\section{Conclusion}\label{conclusions} 
In this work, we studied the guesswork problem under a very general setup of unknown source distribution and decentralized operation. Specifically, we designed and analyzed guessing strategies which do not require the source distribution, the exact guesswork moment to be optimized, or any synchronization between the guesses, yet achieve the optimal guesswork exponent as if all this information was known and full synchronization was possible. Furthermore, we designed efficient algorithms in order to sample guesses from the universal distributions suggested. We believe such sampling methods may be interesting on their own, and find applications outside the guesswork regime.
\section*{Appendix}
\renewcommand{\theequation}{A.\arabic{equation}}
    \setcounter{equation}{0}

\noindent
{\it Proof of Lemma \ref{geometricsum}.}
We denote
\begin{equation}
S=\sum_{k=1}^\infty k^\rho (1-e^{-na})^{k-1}.
\end{equation}
For a given, arbitrarily small $\epsilon > 0$, we first decompose $S$ as
follows.
\begin{equation}
S= \sum_{k=1}^{e^{n(a+\epsilon)}}k^\rho (1-e^{-na})^{k-1}+
\sum_{k=e^{n(a+\epsilon)}+1}^\infty k^\rho (1-e^{-na})^{k-1}\dfn A+B.
\end{equation}
Now,
\begin{eqnarray}
A&\le&\sum_{k=1}^{e^{n(a+\epsilon)}}e^{n(a+\epsilon)\rho}(1-e^{-na})^{k-1}\nonumber\\
&=&e^{n(a+\epsilon)\rho}\sum_{k=1}^{e^{n(a+\epsilon)}}(1-e^{-na})^{k-1}\nonumber\\
&\le&e^{n(a+\epsilon)\rho}\sum_{k=1}^{\infty}(1-e^{-na})^{k-1}\nonumber\\
&=&e^{n(a+\epsilon)\rho}\cdot\frac{1}{1-(1-e^{-na})}\nonumber\\
&=&e^{na}\cdot e^{n(a+\epsilon)\rho}\nonumber\\
&\le&e^{n(1+\rho)(a+\epsilon)}.
\end{eqnarray}
It remains to show then that $B$ has a negligible contribution for large
enough $n$. Indeed, we next show that $B$ decays double--exponentially rapidly in $n$ for
every $\epsilon > 0$.
\begin{eqnarray}
B&=&\sum_{k=e^{n(a+\epsilon)}+1}^\infty k^\rho
(1-e^{-na})^{k-1}\nonumber\\
&=&\sum_{k=e^{n(a+\epsilon)}+1}^\infty \exp\{(k-1)\ln(1-e^{-na})+\rho\ln
k\}\nonumber\\
&=&\sum_{k=e^{n(a+\epsilon)}}^\infty \exp\{k\ln(1-e^{-na})+\rho\ln
(k+1)\}\nonumber\\
&\le&\sum_{k=e^{n(a+\epsilon)}}^\infty \exp\{-k\cdot e^{-na}+\rho\ln
(k+1)\}\nonumber\\
&=&\sum_{k=e^{n(a+\epsilon)}}^\infty
\exp\left\{-k\cdot\left[e^{-na}-\frac{\rho\ln
(k+1)}{k}\right]\right\}
\end{eqnarray}
Since $\{[\ln(k+1)]/k\}_{k\ge 1}$ is a monotonically decreasing sequence, then
for all $k\ge e^{n(a+\epsilon)}$,
$$\frac{\rho\ln(k+1)}{k}\le
\frac{\rho\ln[e^{n(a+\epsilon)}+1]}{e^{n(a+\epsilon)}}=
\rho e^{-n(a+\epsilon)}\ln[e^{n(a+\epsilon)}+1].$$
Thus,
\begin{eqnarray}
B&\le&\sum_{k=e^{n(a+\epsilon)}}^\infty
\exp\{-k\cdot(e^{-na}-\rho
e^{-n(a+\epsilon)}\ln[e^{n(a+\epsilon)}+1])\}\nonumber\\
&=&\frac{\exp\{-e^{n(a+\epsilon)}(e^{-na}-\rho
e^{-n(a+\epsilon)}\ln[e^{n(a+\epsilon)}+1])\}}
{1-\exp\{-(e^{-na}-\rho
e^{-n(a+\epsilon)}\ln[e^{n(a+\epsilon)}+1])\}}\nonumber\\
&=&\frac{\exp\{-(e^{n\epsilon}-\rho\ln[e^{n(a+\epsilon)}+1])\}}
{1-\exp\{-(e^{-na}-\rho
e^{-n(a+\epsilon)}\ln[e^{n(a+\epsilon)}+1])\}}\nonumber\\
&=&\frac{[e^{n(a+\epsilon)}+1]^\rho\exp\{-e^{n\epsilon}\}}
{1-\exp\{-(e^{-na}-\rho
e^{-n(a+\epsilon)}\ln[e^{n(a+\epsilon)}+1])\}}
\end{eqnarray}
Now, for small $x$, we have $1-e^{-x}=x+O(x^2)$, and so, the factor
$$[1-\exp\{-(e^{-na}-\rho
e^{-n(a+\epsilon)}\ln[e^{n(a+\epsilon)}+1])\}]^{-1}$$
is of the exponential order of $e^{na}$, which does not affect the double
exponential decay due to the term $e^{-e^{n\epsilon}}$. The proof of the lemma
is completed by taking into account the arbitrariness of $\epsilon > 0$ (in
particular, one may let $\epsilon$ decay sufficiently slowly with $n$).

\end{document}